\documentclass[aps,prl, showpacs]{revtex4}%
\usepackage{amsfonts}
\usepackage[T2A]{fontenc}
\usepackage[cp1251]{inputenc}

\usepackage{amsmath}
\usepackage{amssymb}
\usepackage[russian,english]{babel}
\usepackage{graphicx}%
\begin{document}
\title{Thermal Hall-Senftleben Effect}
\author{L. A. Maksimov}
\author{T. V. Khabarova}
\email{frau_sych@mail.ru}
\affiliation{Kurchatov Institute, Moscow
123182, Russia}
\date{\today}

\begin{abstract}
This paper is devoted to a prediction of new effect - the occurrence
of a heat flow perpendicular both to temperature gradient and
magnetic field in molecular dielectric where rotary degrees of
freedom of molecules are defrozen. The method of the moments
considering processes of phonon scattering on molecules with their
rotary condition changing is developed.
\end{abstract}

\maketitle

\bigskip






\textbf{1.INTRODUCTION.}

Recently there was a message \cite{Strohm2005}, \cite{Inushkin}\
about detection at presence of a magnetic field in ionic dielectric
$Tb_{3}Ga_{5}O_{12}$ a heat flow perpendicular both to temperature
gradient and magnetic field. This phenomenon is related to
Righi-Leduc effect in metals where it is also observed a component
of a heat flow, directed on $\left[  B,\nabla T\right] $. As in
dielectrics there are no free charge carriers the physical reason of
the phenomenon is far from Hall effect and Righi-Leduc effect
mechanisms. In works \cite{Strohm2005}, \cite{Inushkin} and
theoretical works \cite{Sheng}, \cite{Condmatt} the observable
effect communicates from spin-orbital phonon interaction with
magnetic biased spins of paramagnetic ions. However, there is
another mechanism which can lead to occurrence of transverse heat
flow. In molecular gases this is an anisotropic scattering of
molecules which rotary moments precess in external magnetic field
$\vec{B}$. The observation of a heat flow perpendicular both to
temperature gradient and magnetic field in molecular gases is called
odd Senftleben-Beenakker effect \cite{Beenakker}. In the present
work this effect is generalized on a case of molecular crystals
where oscillating near lattice points molecules can rotate freely at
the same time if temperature is above the rotary degrees of freedom
freezing temperature \cite{Unkn}. Thus as translational motion of
molecules in a solid body is absent, there is no direct analogue of
classical Senftleben-Beenakker effect. Phonons transfer energy in
dielectrics. In a considered solid body we assume there are no
paramagnetic particles and spin-orbital mechanism of a magnetic
field influence on heat conductivity is absent. However, if the
molecule rotary moments of this body can rotate freely in a magnetic
field these moments precess. Precession affects the probability of
anisotropic phonon scattering at nonspherical molecules. It should
lead to phonon energy flow change appeared in presence of
temperature gradient and to occurrence of magnetic field dependence
of thermal conductivity tensor:
\begin{equation}
\varkappa_{ik}(\vec{B})=\varkappa_{ki}(-\vec{B}).
\end{equation}

In the present work it is predicted, that the thermal Hall effect
can be observed not only in ionic, but also in molecular dielectrics
in the field of temperatures where rotary degrees of molecules are
defrozen. Rotary moments $\vec{M}$ will be considered as classical
vectors. (More realistic case of quantum rotation will be
investigated separately). The considered phenomenon is caused by
correlation of phonon movement and the rotary moments and described
by joint distribution function of phonon and rotary degrees of
freedom $f\left( \vec{p},\vec{M}\right)$. Correlation exists, if
phonon scattering at the rotary moments is anisotropic. For an
estimation of such anisotropy it is necessary to solve a mechanical
problem about interaction of two oscillating near adjacent lattice
points molecules. We shall not do this, and using a method of the
moments for solution of the corresponding kinetic equation the
eigenvalues of the collision operator we shall consider as
adjustable parameters. Thus, three parameters are entered into
theories instead of one usual adjustable parameter (phonon
relaxation time $\tau$): 1) reciprocal phonon lifetime owing to
anharmonicity or dispersion on defects  $\Omega_{0}=1/\tau$, 2)
phonon collision frequency with change of a rotary condition of
molecules $\Omega_{1}=\lambda\Omega_{0}$ \ ($\lambda\ll1$ -
nonsphericalness parameter) and 3) precession frequency $\omega
_{B}=-\gamma B$. ($\gamma$ - gyromagnetic ratio). It is used
spherical coordinates and system of units with $\hbar=1,\ k_{B}=1$
here. It will be shown, that the antisymmetric part of heat
conduction tensor is proportional to a magnetic field
\begin{equation}
\varkappa_{ik}^{odd}(\vec{B})\sim\lambda^{2}\left(  \omega_{B}%
/\Omega\right)  .
\end{equation}

\textbf{2.THE KINETIC EQUATION.}

So, we shall find a magnetic field influence on heat conduction of a
molecular crystal with rotary degrees of freedom. We investigate a
simple model of substance in which all molecules or their part have
the rotary moments. We shall consider system as a solution of Debye
phonon gas with a spectrum $\omega_{p}=cp$, and located in units of
lattice of two-atom molecules classical rotary moments $\vec{M}$
noninerating with each other. (In contrast to a problem of heat
conduction in a paramagnetic with spin-phonon interaction
\cite{Condmatt} phonon polarization does not play a role and
detailed elaboration of acoustic phonon properties does not matter).
Phonon and rotary moments interaction is not enough, and in gas
approximation we shall describe evolution of system by Boltzmann
equation \cite{Landau} for joint distribution function $f\left(
\vec{p},\vec{M},\vec{r}\right)  $
\begin{equation}
\frac{\partial f}{\partial t}+\left(  \vec{v}\frac{\partial
f}{\partial\vec
{r}}\right)  +\frac{\partial\vec{r}_{M}}{\partial t}\frac{\partial f}%
{\partial\vec{r}_{M}}+\frac{\partial\vec{M}}{\partial
t}\frac{\partial
f}{\partial\vec{M}}+Stf=0. \label{01}%
\end{equation}
Here $\vec{r},\vec{v}=\partial\omega/\partial\vec{p}\ $ are a phonon
coordinate and velocity ($\left\vert \vec{v}\right\vert =c$). If the
subsystem of rotary moments could move like gas in space, it would
be necessary to bring in the left part of
(\ref{01}) an expression $\frac{\partial\vec{r}_{M}}{\partial t}%
\frac{\partial f}{\partial\vec{r}_{M}}$, where $\vec{r}_{M}$ is a
coordinate of a point where one of the rotary moments is located.
But such motion is absent and $\frac{\partial\vec{r}_{M}}{\partial
t}\equiv0$. The fourth term at the left describes rotary moment
precession in a magnetic field $\left(  \frac{\partial }{\partial
t}\vec{M}=\gamma\lbrack\vec{M}\vec{B}]\right)  $. In paramagnetic
substance, there is a mix of components with different electron spin
directions. The odd effect is created by component, consisting of
molecules which have a spin projection on molecule axis equal to
zero. In nonparamagnetic body molecule gyromagnetic ratio $\gamma$\
is much less, but not zero because of nuclei slipping in relation to
electron shell \cite{Ramsey}. For simplicity we shall consider a
case of nonparamagnetic molecules. Below we shall generally use
spherical coordinate representation where precession operator has as
much simple form as possible:
\begin{equation}
\hat{\Omega}_{B}=\gamma\lbrack\overrightarrow{M}\overrightarrow{B}%
]\frac{\partial}{\partial\overrightarrow{M}}=\omega_{B}\frac{\partial
}{\partial\varphi_{M}},\ \omega_{B}=-\gamma B,\ \ \hat{\Omega}_{B}%
Y_{l_{2}m_{2}}(\vec{M})=im_{2}\omega_{B}Y_{l_{2}m_{2}}(\vec{M}) \label{42}%
\end{equation}
In a state of thermodynamic equilibrium we have
\begin{equation}
f^{(0)}\left(  \vec{p},\vec{M}\right)
=N^{(0)}(\vec{p})\varphi^{(0)}(\vec
{M}), \label{21}%
\end{equation}
where
\begin{equation}
N^{(0)}(\vec{p})=(\exp(\omega_{p}/T)-1)^{-1},\ \ \varphi^{(0)}(\vec{M}%
)=\frac{1}{4\pi IT}\exp(-\varepsilon/T) \label{22}%
\end{equation}
($\varepsilon=M^{2}/2I$ and $I$\ are energy and the moment of
inertia of a two-atom molecule). The phonon number is set by
temperature, and the rotary moment number is fixed. Distribution
function ($\varphi^{(0)}(\vec{M})$) is referred to one molecule. At
the decision of a stationary problem of heat conductivity the first
term in (\ref{01}) is absent, and the second is equal to
\begin{equation}
\left(  \vec{v}\nabla\right)  f^{(0)}=\left(  \vec{v}\nabla
N^{(0)}(\vec {p},\vec{r})\right)  \varphi^{(0)}=\left( \vec{v}\nabla
T\right) (\omega/T^{2})N^{(0)}\left( N^{(0)}+1\right)
\varphi^{(0)}=f^{(0)}\vec
{Q}\nabla T, \ \  \vec{Q}=\vec{v}\frac{\omega}{T^{2}}(\left(  N^{(0)}%
+1\right)  \label{100}%
\end{equation}
in linear approximation on a temperature gradient. The rotary
moments are attached to a lattice and cannot give the contribution
to a heat flow. Therefore calculating a gradient we have neglected
spatial heterogeneity.

Let's represent distribution function in the form of
\begin{equation}
f=f^{(0)}(1+\vec{\chi}\nabla T) \label{1}%
\end{equation}

In linear approximation on $\nabla T$ the collision integral is
expressed through the linear integral operator
\begin{equation}
Stf=\left(  \nabla T\right)  f^{(0)}\hat{\Omega}\vec{\chi}, \label{02}%
\end{equation}
which acts on $\vec{\chi}$. Positive definite Hermitian collision
operator $\hat{\Omega}$ is completely described by the matrix
elements $\Omega_{ab}=\left\langle
\psi_{a}^{\ast}\hat{\Omega}\psi_{b}\right\rangle $. Matrix elements
$\Omega_{ab}$ possess properties:
\begin{equation}
\left\langle \psi_{a}^{\ast}\hat{\Omega}\psi_{b}\right\rangle ^{\ast
}=\left\langle \psi_{b}^{\ast}\hat{\Omega}\psi_{a}\right\rangle
=\left\langle \left(  \hat{\Omega}\psi_{a}\right)
\psi_{b}^{\ast}\right\rangle ,\ \ \left\langle
\psi_{a}^{\ast}\hat{\Omega}\psi_{a}\right\rangle \geq0
\label{11}%
\end{equation}
As it has been already noted in introduction, we shall not write
explicit rotary moment dependence of phonon scattering cross-section
and we shall apply standard approach in kinetics when dispersion
properties are described phenomenologically by $\Omega_{ab}$ which
are assumed known, i.e., actually, they are adjustable parameters.
Thus, removing from (\ref{01}) a time derivative, using
(\ref{100}),(\ref{42}),(\ref{02}), and reducing on $f^{(0)}\nabla
T$, we obtain the vector integro-differential equation.
\begin{equation}
(\hat{\Omega}+\hat{\Omega}_{B})\vec{\chi}+\vec{Q}=0, \label{3}%
\end{equation}

\textbf{3. METHOD OF MOMENTS.}

Value of a heat flow and heat conduction tensor in Debye model are
defined by
\begin{equation}
q_{i}=%
{\displaystyle\sum}
(v_{i}\omega)f=\int\frac{d^{3}p}{\left(  2\pi\right)  ^{3}}\frac{d^{3}M}%
{M}\left(  v_{i}\omega\right)  \left(  f^{(0)}\vec{\chi}\nabla
T\right) =-\varkappa_{ik}\left(  \nabla T\right)  _{k},\
\varkappa_{ik}=-\left\langle
v_{i}\omega\chi_{k}\right\rangle , \label{6}%
\end{equation}
where $\left\langle A\right\rangle \equiv\int\frac{d^{3}p}{\left(
2\pi\right) ^{3}}\frac{d^{3}M}{M}f^{(0)}A$.

So, the problem comes to the solving of vector equation (\ref{3}).
Vector-function $\vec{\chi}$ we shall search in the form of
expansion
\begin{equation}
\chi_{i}=\sum_{a}\chi_{ia}\psi_{a}\left(  \vec{p},\vec{M}\right)  \label{7}%
\end{equation}
on orthonormal vector functions $\psi_{a}\left(  \vec{p},\vec
{M}\right)  :\left\langle \psi_{a}^{\ast}\psi_{b}\right\rangle
=\delta_{ab}$. Orthonormal functions $\psi_{a}$ are convenient for
choosing in the form of superposition of products of spherical
polynoms depending on impulse directions $Y_{l_{1}m_{1}}%
(\vec{p})$ and rotary moment $Y_{l_{2}m_{2}}(\vec{M})$
\begin{equation}
\psi_{1ml_{1}l_{2}}\sim\sum_{m_{1}m_{2}}C_{l_{1}m_{1}l_{2}m_{2}}^{1m}%
Y_{l_{1}m_{1}}(\vec{p})Y_{l_{2}m_{2}}(\vec{M}), \label{90}%
\end{equation}
where $C_{l_{1}m_{1}l_{2}m_{2}}^{1m}$ is Clebsch-Gordan
coefficients. To simplify notations, we do not write explicitly
normalization and factors containing scalar polynoms depending on
translational and rotary energies. Transition from cartesian
coordinates to spherical is carried out by coefficients $t_{im}$,
where $t_{xm}=\sqrt{\frac{2\pi}{3}},t_{ym}=\frac{m}{i}\sqrt{\frac{2\pi}{3}}%
,m=\pm1$
\begin{equation}
Q_{i}=\sum_{m}t_{im}Q_{1m},\ \ \varkappa_{ik}=-\left\langle
v_{i}\omega \chi_{k}\right\rangle
=-\sum_{m}t_{im}^{\ast}t_{km}\left\langle v_{1m}^{\ast
}\omega\chi_{1m}\right\rangle
\end{equation}
In particular, off-diagonal element of heat conduction tensor
$\varkappa_{yx}$ is proportional to $\operatorname{Im}\left\langle
v_{11}^{\ast}\omega \chi_{11}\right\rangle $. It is responsible for
a heat flow in a direction (y), perpendicular to temperature
gradient (x) and magnetic field (z). For solving equation (\ref{3})
in 2-moment approximation as a first function from (\ref{7}) we
shall choose wave function contained in $\overrightarrow {Q}$, which
describes heterogeneity in (\ref{3}), i.e.
\begin{equation}
\psi_{1m10}=v_{1m}\omega\label{71}%
\end{equation}
As a second function we shall choose vector-function which is even
on $\vec{M}$ and odd on $\vec{p}$. The simplest expression,
satisfying to this requirement in cartesian coordinates is
$\vec{M}\left(  \vec{M} \vec{p}\right)  $, or in spherical
coordinates
\begin{equation}
\psi_{1m12}=\sum_{m_{1}m_{2}}C_{1m_{1}2m_{2}}^{11}Y_{1m_{1}}(\vec{p}%
)Y_{2m_{2}}(\vec{M}). \label{72}%
\end{equation}

Term $\operatorname{Im}\left\langle
v_{11}^{\ast}\omega\chi_{11}\right\rangle $ is easy to find, if the
probability of molecula rotary condition changing is small (weak
nonsphericity) at phonon dispersion.

Let in zeroth-order approximation the collision operator in
representation of functions (\ref{90}) be reduced by one relaxation
time
\begin{equation}
\hat{\Omega}^{(0)}\psi_{n}=\Omega_{0}\psi_{n},\ \ \Omega_{0}=1/\tau,
\label{10}%
\end{equation}
and off-diagonal part $\lambda\hat{\Omega}^{(1)}$, proportional to
small parameter $\left(  \lambda\ll1\right)$, will transform the
functions depending only from an impulse, to functions of a kind
(\ref{90}), depending from $\vec{M}$
\begin{equation}
\hat{\Omega}^{(1)}\psi_{1m10}=\Omega_{1}\psi_{1m12} \label{121}%
\end{equation}
The equation (\ref{3})
\begin{equation}
(\hat{\Omega}^{(0)}+\lambda\hat{\Omega}^{(1)}+\hat{\Omega}_{B})\chi
_{11}+Q_{11}=0\label{8}%
\end{equation}
is solved by decomposition of function $\vec{\chi}$ on degrees of
$\lambda$. We shall determine operator $\hat{K}$ by expression
\begin{equation}
\hat{K}=(\hat{\Omega}^{(0)}+\hat{\Omega}_{B})^{-1}%
\end{equation}
The equation (\ref{8}) we shall write down in the form of
\begin{equation}
(\hat{K}^{-1}+\lambda\hat{\Omega}^{(1)})\chi_{11}+Q_{11}=0
\end{equation}
Then we shall increase it on $\hat{K}$ at the left and then - on
$(1+\lambda\hat{K}\hat{\Omega} ^{(1)})^{-1}$. From here we find
\begin{equation}
\chi_{11}=-(1+\lambda\hat{K}\hat{\Omega}^{(1)})^{-1}\hat{K}Q_{11}%
=-(1-\lambda\hat{K}\hat{\Omega}^{(1)}+\lambda^{2}\hat{K}\hat{\Omega}^{(1)}%
\hat{K}\hat{\Omega}^{(1)}+...)\hat{K}Q_{11}%
\end{equation}
The required term is equal to
\begin{equation}
\operatorname{Im}\left\langle
v_{11}^{\ast}\omega\chi_{11}\right\rangle
=-\operatorname{Im}\left\langle v_{11}^{\ast}\omega(1-\lambda\hat{K}%
\hat{\Omega}^{(1)}+\lambda^{2}\hat{K}\hat{\Omega}^{(1)}\hat{K}\hat{\Omega
}^{(1)}+...)\hat{K}Q_{11}\right\rangle \label{9}%
\end{equation}
As function $Q_{11}$ does not depend from $\vec{M}$ operator $\hat
{\Omega}_{B}$ does not act on it, and
\begin{equation}
\hat{K}Q_{11}=(\hat{\Omega}^{(0)})^{-1}Q_{11}=Q_{11}/\Omega_{0}.
\end{equation}
As a result the right side (\ref{9}) becomes
\begin{equation}
-\operatorname{Im}\left\langle v_{11}^{\ast}\omega(1-\lambda\frac{1}%
{\Omega_{0}}\hat{\Omega}^{(1)}+\lambda^{2}\frac{1}{\Omega_{0}}\hat{\Omega
}^{(1)}\hat{K}\hat{\Omega}^{(1)}+...)Q_{11}\right\rangle \frac{1}{\Omega_{0}}%
\end{equation}
First two terms give zero. The nonvanishing contribution gives a
term of the second order on $\lambda$, and accurate within numerical
factor the off-diagonal part of heat conduction tensor is equal to
\begin{equation}
\varkappa_{yx}\sim\frac{\lambda^{2}}{\Omega_{0}^{2}T^{2}%
}\operatorname{Im}\left\langle \left(  v_{11}^{\ast}\omega\right)
\hat {\Omega}^{(1)}\hat{K}\hat{\Omega}^{(1)}\left(
v_{11}\omega\right)  \left(
N^{(0)}+1\right)  \right\rangle \label{98}%
\end{equation}
As the given frequencies integral quickly converges both at low and
at high frequencies, then it is possible to neglect $\left(
N^{(0)}+1\right)$ and write in a compact kind
\begin{equation}
\varkappa_{yx}\sim\frac{\lambda^{2}}{\Omega_{0}^{2}T^{2}%
}\operatorname{Im}\left\langle P^{\ast}\hat{K}P\right\rangle ,
\end{equation}
where function $P=\hat{\Omega}^{(1)}\left(  v_{11} \omega\right)$ is
introduced. In concordance with (\ref{71}) and (\ref{121}) it is
written in form of
\begin{equation}
P=\Omega_{1}\psi_{1m12}.
\end{equation}
To calculate a matrix element $\left\langle
P^{\ast}\hat{K}P\right\rangle $ it is necessary to find function
\begin{equation}
F=\hat{K}P=(\hat{\Omega}^{(0)}+\hat{\Omega}_{B})^{-1}P
\end{equation}
For this purpose we should solve following equation
\begin{equation}
\left(  \hat{\Omega}^{(0)}+\omega_{B}\frac{\partial}{\partial\varphi_{M}%
}\right)
F=\Omega_{1}\sum_{m_{1}m_{2}}C_{1m_{1}2m_{2}}^{11}Y_{1m_{1}}(\vec
{p})Y_{2m_{2}}(\vec{M}),
\end{equation}
The solution will be searched in the form of
\begin{equation}
F=\sum_{m_{1}m_{2}}F_{m_{2}}C_{1m_{1}2m_{2}}^{11}Y_{1m_{1}}(\vec{p})Y_{2m_{2}%
}(\vec{M}).
\end{equation}
Expansion coefficients are easy to be found in approximation of one
relaxation time: $\hat{\Omega}^{(0)}F=\Omega_{0}F$. Then
\begin{equation}
\left(
\Omega_{0}+\omega_{B}\frac{\partial}{\partial\varphi_{M}}\right)
F=\sum_{m_{1}m_{2}}\left(  \Omega_{0}+im_{2}\omega_{B}\right)  F_{m_{2}%
}C_{1m_{1}2m_{2}}^{11}Y_{1m_{1}}(\vec{p})Y_{2m_{2}}(\vec{M})=\Omega_{1}%
\sum_{m_{1}m_{2}}C_{1m_{1}2m_{2}}^{11}Y_{1m_{1}}(\vec{p})Y_{2m_{2}}(\vec{M})
\end{equation}
From here we obtain
\begin{equation}
F_{m_{2}}=\Omega_{1}\left(  \Omega_{0}+im_{2}\omega_{B}\right)
^{-1},
\end{equation}
and the value interesting us (\ref{11}) becomes
\begin{equation}
\varkappa_{yx}\sim\frac{\lambda^{2}}{\Omega_{0}^{2}T^{2}%
}\operatorname{Im}\left\langle \left(  \sum_{m_{1}^{/}m_{2}^{/}}C_{1m_{1}%
^{/}2m_{2}^{/}}^{11}Y_{1m_{1}^{/}}(\vec{p})Y_{2m_{2}^{/}}(\vec{M})\right)
^{\ast}\sum_{m_{1}m_{2}}\Omega_{1}\left(
\Omega_{0}+im_{2}\omega_{B}\right)
^{-1}C_{1m_{1}2m_{2}}^{11}Y_{1m_{1}}(\vec{p})Y_{2m_{2}}(\vec{M})\right\rangle
\label{12}%
\end{equation}
After averaging on $\vec{p}$ and $\vec{M}$ we come to final
expression:
\begin{equation}
\varkappa_{yx}\sim\frac{\lambda^{2}\Omega_{1}}{\Omega_{0}^{2}}%
\sum_{m_{1}m_{2}}\left(  C_{1m_{1}2m_{2}}^{11}\right)  ^{2}\frac{m_{2}%
\omega_{B}}{\Omega_{0}^{2}+\left(  m_{2}\omega_{B}\right)  ^{2}}%
\end{equation}

We see, that the perpendicular heat flow is odd on a field and is
maximal, when order of magnitude of $\xi=\omega_{B}/\Omega_{0}$ is
1.

Here the case of nonparamagnetic molecules was considered. In a
solid body at achievable fields effect seems to be much less than
the maximal value, and the perpendicular heat flow grows linearly
with a field
\begin{equation}
\varkappa_{yx}\sim\lambda^{2}\omega_{B}\Omega_{0}^{-4}%
\end{equation}

The even effect in a magnetic field is described by the valid part
of expression which imaginary part was written in (\ref{12}):
\begin{equation}
\varkappa_{xx}\sim\frac{\lambda^{2}\Omega_{1}}{\Omega_{0}^{2}}%
\sum_{m_{1}m_{2}}\left(  C_{1m_{1}2m_{2}}^{11}\right)  ^{2}\frac{\Omega_{0}%
}{\Omega_{0}^{2}+\left(  m_{2}\omega_{B}\right)  ^{2}}%
\end{equation}

Even and odd effects for gas were experimentally observed many times
(see for example \cite{Beenakker} and \cite{Ref8}). For a solid body
with rotary degrees of freedom the of magnetic field influence on
heat conductivity was not experimentally researched yet.

In the present work it is shown, that in molecular crystals where
quasi-free molecule rotation exists in wide range of temperatures,
the heat conductivity effect, caused by occurrence of a  heat flow
perpendicular to a temperature gradient, similar to the Hall effect
in metals should be observed odd on a magnetic field.

\end{document}